\documentclass{article}

\usepackage{arxiv}

\usepackage[utf8]{inputenc} 
\usepackage[T1]{fontenc}    
\usepackage{hyperref}       
\usepackage{url}            
\usepackage{booktabs}       
\usepackage{amsfonts}       
\usepackage{nicefrac}       
\usepackage{microtype}      
\usepackage{lipsum}		
\usepackage{microtype,mathtools}
\usepackage{amsmath,amssymb,amsthm}
\usepackage{graphicx}
\usepackage{natbib}
\let\cite=\citep

\title{Realistic Dune Field Surface Stress Prediction (Technical Report)}

\author{
  Chao Wang, Ph.D.\\
  The University of Texas at Dallas\\
  Richardson, TX\\
  \texttt{cxw151530@utdallas.edu} \\
}

\begin{document}
\maketitle

\begin{abstract}
The dune morphodynamics study is under highly focused recently, due to aeolian process induced nonlinear correlation to sediment modification over bedform. Surface stress, inflicted by aloft upcoming wind, impacts, crucially, the sediment erosion pattern. The aloft atmospheric dune surface layer (ASL) is composed of inertial layer and roughness sublayer. The logarithmic velocity profile is valid within inertial layer based on previous findings, where attached-eddy occupied the higher elevations scaled with local wall-normal height \cite{wang2019turbulence}. While, mixing layer velocity profile is displayed in this work to predict the surface shear over White Sands National Monument (WSNM), the length scale of which is revealed as proportional to mixing layer length scale. Given the realistic dune field morphology and aeolian process, a series of mathematic stress models have been proposed and verified. Among these models, mixing layer model shows the best performance, which elucidates and confirms the mixing layer analogy over realistic dune field. Finally, a schematic structural model shows that ``sediment scour'' and ``flow channeling'' indeed is the reason for the large residual values distributed in narrow interdune surfaces, which has a great consistency with previous finding \cite{wang2018large}.
\end{abstract}

\section{Large-Eddy Simulation \& Cases}
\label{s:les_case}

Recently, Computational Fluid Dynamics (CFD) method benefits a wide range of researches such as biomedicine, environment, geoscience and so forth, wherein different length scales of fluid are involved in these CFD simulations, especially in turbulent flow which consisted of a broad range of spectrum \citep{wang2019thesis}. To recover the geophysical fluid field, Direct Numerical Simulation (DNS) could be a promising methodology, resolving turbulence from Kolmogorov scale to the maximum of numerical length scale \citep{pope00}. However, DNS is too ``expensive'' for most geophysical fluid simulations due to the critical computational resource need in high Reynolds number flow regime. In contrast, LES can approach highly credible results with an acceptable numerical requirement \citep{metias96}. In this chapter, Large-Eddy Simulation method will be detailed discussed. Meanwhile, in Section \ref{sb:cases}, all numerical cases details will be provided. 

\subsection{Large-Eddy Simulation}
\label{sb:les}
In Large-Eddy Simulation (LES) method, the filtered three-dimensional transport equation, incompressible momentum, 
\begin{equation}
D_t \tilde{\boldsymbol u} ({\boldsymbol x},t) = \rho^{-1} {\boldsymbol F}({\boldsymbol x},t), 
\end{equation}
is solved, where $\rho$ is density, $\tilde{.}$ denotes a grid-filtered quantity, $\boldsymbol{u}(\boldsymbol{x},t)$ is velocity (in this work, $u$, $v$, $w$ are corresponded to velocity in streamwise, spanwise and wall-normal direction, respectively) and $\boldsymbol{F}(\boldsymbol{x},t)$ is the collection of forces (pressure correction, pressure gradient, stress heterogeneity and obstacle forces). The grid-filtering operation is attained here via convolution with the spatial filtering kernel, $\tilde{\boldsymbol u} ({\boldsymbol x},t) = G_\Delta \star {\boldsymbol u} ({\boldsymbol x},t)$, or in the following form
\begin{equation}
\tilde{\boldsymbol{u}}(\boldsymbol{x},t)=\oint G_{\Delta}(\boldsymbol{x}-\boldsymbol{x}^{\prime},t)\boldsymbol{u}(\boldsymbol{x}^{\prime},t)\mathrm{d}\boldsymbol{x}^{\prime},
\end{equation}
where $\Delta$ is the filter scale \citep{meneveaukatz}. A right-hand side forcing term, $- \nabla \cdot \boldsymbol{\mathsf{T}}$, will be generated after the filtering operation to momentum equation, where $\boldsymbol{\mathsf{T}} = \langle {\boldsymbol u}^\prime \otimes {\boldsymbol u}^\prime \rangle_t$ is the subgrid-scale stress tensor and $\langle . \rangle_a$ denotes averaging over dimension, $a$ (in this article, rank-1 and -2 tensors are denoted with bold-italic and bold-sans relief, respectively).

For the present study, $D_t \tilde{\boldsymbol u} ({\boldsymbol x},t) = \rho^{-1} {\boldsymbol F} ({\boldsymbol x},t)$ is solved for a channel-flow arrangement \citep{albertsonparlange1999,AndersonChamecki14}, with the flow forced by a pressure gradient in streamwise direction, $\boldsymbol{\mathrm{\Pi}} = \{\mathrm{\Pi},0,0\}$, where 
\begin{equation}
\mathrm{\Pi} = \left[ \mathrm{d}P_{0}/\mathrm{d}x\right]\frac{H}{\rho}=\tau^{w}/\rho=u_{*}^{2}=1,
\end{equation}
which sets the shear velocity, $u_{*}$, upon which all velocities are non-dimensionalized. In simulation, all length scales are normalized by $H$, which is the surface layer depth, and velocity are normalized by surface shear velocity. $D_t \tilde{\boldsymbol u} ({\boldsymbol x},t) = \rho^{-1} {\boldsymbol F} ({\boldsymbol x},t)$ is solved for high-Reynolds number, fully-rough conditions \citep{Jimenez2004}, and thus viscous effects can be neglected in simulation, $\nu \nabla^2 \tilde{\boldsymbol u} ({\boldsymbol x},t) = 0$. Under the presumption of $\rho ({\boldsymbol x},t) \rightarrow \rho$, the velocity vector is solenoidal, $\nabla \cdot \tilde{\boldsymbol u} ({\boldsymbol x},t) = 0$. During LES, the (dynamic) pressure needed to preserve $\nabla \cdot \tilde{\boldsymbol u} ({\boldsymbol x},t) = 0$ is dynamically computed by computation of $\nabla \cdot \left[ D_t \tilde{\boldsymbol u} ({\boldsymbol x},t) = \rho^{-1} {\boldsymbol F} ({\boldsymbol x},t) \right]$ and imposing $\nabla \cdot \tilde{\boldsymbol u} ({\boldsymbol x},t) = 0$, which yields a resultant pressure Poisson equation. 

The channel-flow configuration is created by the aforementioned pressure-gradient forcing, and the following boundary condition prescription: at the domain top, the zero-stress Neumann boundary condition is imposed on streamwise and spanwise velocity, $\partial \tilde{u} / \partial z | _{z / H = 1} = \partial \tilde{v} / \partial z | _{z / H = 1} = 0$. The zero vertical velocity condition is imposed at the domain top and bottom, $\tilde{w} (x , y , z / H = 0) = \tilde{w} (x,y,z / H = 1) = 0$. Spectral discretization is used in the horizontal directions, thus imposing periodic boundary conditions on the vertical ``faces'' of the domain, \textit{vis.}
\begin{equation}
\phi(x+mL_x,y+nL_y,z)=\phi(x,y,z), 
\end{equation}
and imposing spatial homogeneity in the horizontal dimensions. The code uses a staggered-grid formulation \citep{albertsonparlange1999}, where the first grid points for $\tilde{u}({\boldsymbol x},t)$ and $\tilde{v}({\boldsymbol x},t)$ are located at $\delta z/2$, where $\delta z = H/N_z$ is the resolution of the computational mesh in the vertical ($N_z$ is the number of vertical grid points). Grid resolution in the streamwise and spanwise direction is $\delta x = L_x/N_x$ and $\delta y = L_y/N_y$, respectively, where $L$ and $N$ denote horizontal domain extent and corresponding number of grid points (subscript $x$ or $y$ denotes streamwise or spanwise direction, respectively). Table 1 provides a summary of the domain attributes for the different cases, where the domain height has been set to the depth of the surface layer, $L_z/H = 1$.

At the lower boundary, surface momentum fluxes are prescribed with a hybrid scheme leveraging an immersed-boundary method (IBM)\citep{AndersonMeneveau2010,Anderson12} and the equilibrium logarithmic model \citep{piomelli} , depending on the digital elevation map, $h(x,y)$. When $h(x,y) < \delta z/2$, the topography vertically unresolved, and the logarithmic law is used:
\begin{equation} \label{Equation1}
\tau_{xz}^{w}(x,y,t)=-\left[{\frac{{\kappa}U(x,y,t)}{\log(\frac{1}{2}\delta{z}/\hat{z}_{0})}}\right]^{2}\frac{\bar{\tilde{u}}(x,y,\frac{1}{2}\delta{z},t)}{U(x,y,t)}
\end{equation}
and
\begin{equation} \label{Equation2}
\tau_{yz}^{w}(x,y,t)=-\left[{\frac{{\kappa}U(x,y,t)}{\log(\frac{1}{2}\delta{z}/\hat{z}_{0})}}\right]^{2}\frac{\bar{\tilde{v}}(x,y,\frac{1}{2}\delta{z},t)}{U(x,y,t)}
\end{equation}
where $\hat{z}_{0}/H=2\times10^{-4}$ is a prescribed roughness length, $\bar{\tilde{.}}$ denotes test-filtering \citep{germanoJFM,germanoetal} (used here to attenuate un-physical local surface stress fluctuations associated with localized application of Equation \ref{Equation1} and \ref{Equation2} \citep{bouzeidetal2005a}), and $U(x,y,\frac{1}{2}\delta{z},t)=(\bar{\tilde{u}}(x,y,\frac{1}{2}\delta{z},t)^{2}+\bar{\tilde{v}}(x,y,\frac{1}{2}\delta{z},t)^{2})^{1/2}$ is magnitude of the test-filtered velocity vector. Where $h(x,y) > \frac{1}{2}\delta{z}$, a continuous forcing Iboldsymbol is used \citep{Anderson12,mittaliaccarino2005}, which has been successfully used in similar studies of turbulent obstructed shear flows \citep{AndersonChamecki14,Andersonetal15b,anderson16}. The immersed boundary method computes a body force, which imposes circumferential momentum fluxes at computational ``cut'' cells based on spatial gradients of $h(x,y)$:
\begin{equation}
{\boldsymbol f}({\boldsymbol x},t)=-\frac{\tilde{\boldsymbol{u}}(\boldsymbol{x},t)}{\delta z} R(\tilde{\boldsymbol{u}}(\boldsymbol{x},t)\cdot \nabla h), 
\label{IBM}
\end{equation}
where $R$ is called Ramp Function \citep{Anderson12}
\begin{equation}
R(x)=\begin{cases}
    x & \text{for $x>0$},\\
    0 & \text{for $x\leqslant0$}.
  \end{cases}
\end{equation}
 Equations \ref{Equation1} and \ref{Equation2} are needed to ensure surface stress is imposed when $h(x,y) < \frac{1}{2}\delta{z}$. Subgrid-scale stresses are modeled with an eddy-viscosity model, 
\begin{equation}
{\boldsymbol \tau}^{d}=-2\nu_{t}\boldsymbol{\mathsf{S}}, 
\end{equation}
where 
\begin{equation}
\boldsymbol{\mathsf{S}}=\frac{1}{2}(\nabla {\tilde{\boldsymbol u}}+\nabla {\tilde{\boldsymbol u}}^\mathrm{T}) 
\label{facets}
\end{equation}
is the resolved strain-rate tensor. The eddy viscosity is 
\begin{equation}
\nu_{t}=(C_{s}\Delta)^{2}|\boldsymbol{\mathsf{S}}|, 
\end{equation}
where $|\boldsymbol{\mathsf{S}}|=(2\boldsymbol{\mathsf{S}} \boldsymbol{:} \boldsymbol{\mathsf{S}})^{1/2}$, $C_{s}$ is the Smagorinsky coefficient, and $\Delta$ is the grid resolution. For the present simulations, the Lagrangian scale-dependent dynamic model is used \citep{bouzeidetal2005a}. The simulations have been run for $N_t \delta_t U_0 u_{*,d} H^{-1} \approx 10^3$ large-eddy turnovers, where $U_{0}=\langle{\tilde{u}(x,y,(L_z - \delta z/2)/H=1,t)}\rangle_{t}$ is a ``free stream'' or centerline velocity. This duration is sufficient for computation of Reynolds-averaged quantities. 

\subsection{Cases}
\label{sb:cases}

The White Sands National Monument (WSNM) dune field is located in the Tularosa Basin of the Rio Grande Rift, between the San Andres and Sacramento Mountain Ranges, in souther New Mexico. The WSNM dune field consists of a core of barchan dunes, which abruptly transition to parabolic dunes \cite{ewingkocurek10a,ewingkocurek10b,jerolmackmohrig}. Recently, the increasing trend of aerodynamic roughness of dune field in the upcoming wind direction has been imputed to the developing of the internal momentum boundary layer \cite{jerolmackmohrig}. The WSNM DEM is taken from an existing LiDAR survey. \cite{AndersonChamecki14} has chosen a series windows of WSNM DEM and analyzed DFSL which depicts profound turbulence enveloped beneath shear layers for elevation less than two to three times the dune crest height. For a comprehensive understanding of the turbulence coherence in DFSL, a portion of WSNM DEM has been chosen for this study. Figure 5 (b) displays the subset area of WSNM used as a lower boundary during LES. In terms of geometric complexity, the DEM serves as an `upper limit' with its multiscale distribution of dune sizes and shapes. Importantly, the feature of WSNM reveal the most common and typical dune field which includes overlapping and collision with adjacent dunes, which confounds efforts to isolate universal flow pattern. 

Two \textit{a priori} modifications are imposed on the chosen DEM: \textit{(i)} the lowest elevation is subtracted from the DEM, which imposes the minimum elevation of DEM is $0$, $min(h(\boldsymbol{x}))=0$; \textit{(ii)} a two-dimensioanl windowing function $\mathcal{W}(x,y)=a(x)a(y)$ to the resulting topography in order to impose periodicity on the underlying topography $h(x,y)$. Hence, the modified topography is the Hadamard product $h(x,y)\Rightarrow h(x,y)\mathcal{W}(x,y)$. Periodicity is needed due to the use of spectral decomposition of flow quantities in LES. The Gibbs phenomena will contaminate the results if $h(x,y)$ is not periodic \cite{tsengmeneveau}. The windowing function is in the following format: 
\begin{equation}
a(x)=\begin{cases}
    1.0 & \text{for $0\leqslant\|x-x_c\|<\gamma L$},\\
    cos\left[\frac{(x-x_c)/H-\gamma(L/H)}{2(1-\gamma)}\right] & \text{for $\gamma L\leqslant\|x-x_c\|< L$},
  \end{cases}
  \label{eq:window}
\end{equation}
where $L$ is the length of simulation domain ($L=1000m$ in this work), $H$ is the simulation characteristic length scale to normalize all lengths ($H=100m$ in this work, which is sufficient to recover dune roughness sublayer and inertial sublayer). The parameter $\gamma$ imposes the circumjacent width over which the topography around the edges of the focus area is forced toward $h(x,y) = 0$, and $x_c$ is the coordinate of the center of the domain. In the present study, we select $\gamma = 0.85$, which imposes that the outermost $15\%$ of $h(x,y)$ gradually tends toward zero (see Figure 1 (a)). 

\begin{figure}
\begin{center}
\includegraphics[width=16.5cm]{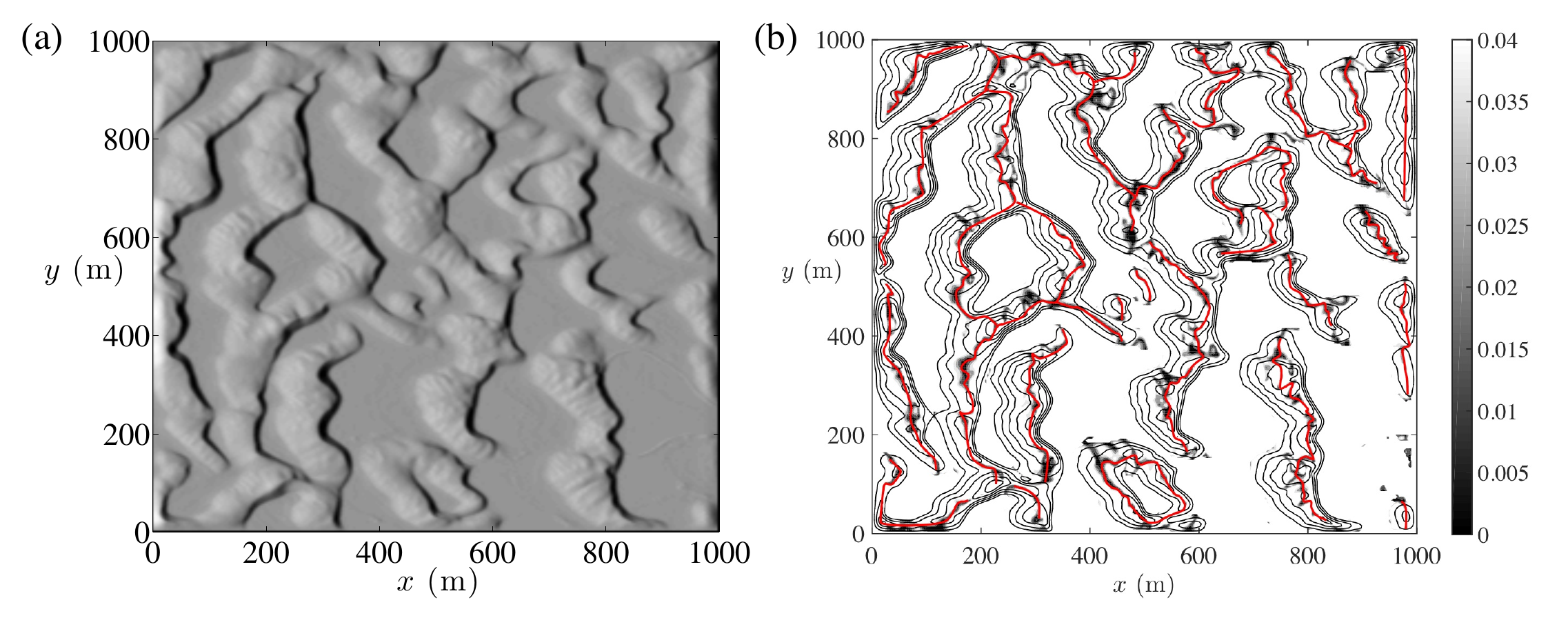}
\caption{Panel (a) is White Sand National Monument Digital Elevation Model (DEM). Solar lighting has been included to highlight the dune field geometry, wherein light gray indicates stoss side, and dark gray indicates lee side. Based on Panel (a), the dune brinklines are captured based on equation \ref{gradient} in Panel (b). The red line marks dune field ridges. The black area indicates ${\partial{h}}/{\partial{x}}\approx 0$ area.}
\label{Figure1}
\end{center}
\end{figure}

Figure 1 (a) displays the White Sands National Monument (WSNM) DEM. Figure 1 (b) captured the brinklines of WSNM via
\begin{equation}
\frac{\partial{h}}{\partial{x}}\approx 0.
\label{gradient}
\end{equation}
The interactive collision is evident in realistic dune field, associated with red lines tangling with each other. From brinkline map, the length scale of brinkline is much larger than the dune height, where the maximum can be over hundred times the crest height. The meandering brinkline pattern is induced by the local non-uniformity of wind regime such as wind direction, velocity magnitude etc. In the following chapter, we will develop a series of stress model, given the known aeolian process models over dune fields, and assess the verification of each model through residual value calculating practices. 

\section{Results}
\label{s:results}

Surface shear stress magnitude over White Sands National Monument has been displayed in Figure 2. The stress magnitude is calculated via IBM (Eq.\ref{IBM}) \cite{Anderson12}. Panel (a) shows the surface shear stress magnitude. The shear stress $\tau_{w,0}(\boldsymbol{x})$ is normalized by the global friction velocity $\tau_{w}(\boldsymbol{x})=\tau_{w,0}(\boldsymbol{x})/u_{*,d}^2$. From Panel (a), we can see the high stress area is on the dune field windward faces. Meanwhile, with the wall-normal elevation increasing, the stress magnitude is enhanced. To elucidate the aeolian stress over dune surface, Panel (b) shows surface shear stress on dunes which excludes the values on walls, $\tau_{w}(x,y,z/h_w>0)$. Panel (c) highlights the windward area stress via $\tau_w(\boldsymbol{x})S(\boldsymbol{x})$, where
\begin{equation}
S(x)=\begin{cases}
    0 & \text{for $\frac{\partial{h}}{\partial{x}}<0$},\\
    1 & \text{for $\frac{\partial{h}}{\partial{x}}>0$}.
  \end{cases}
  \label{Sx}
\end{equation}
In general, the leeward stress magnitude is less than $0.1$. In Panel (c), solid black line on colorbar highlights the critical magnitude on stoss side, $\tau_{w}>0.4$. However, the high stress value can also be captured in some leeward faces and the areas where ``flow-channeling'' appears \cite{wang2017numerical,wang2017large,wang2018large2,wang2018large,wang2019turbulence,wang2019numerical2,wang2019thesis}. 

\begin{figure}
\begin{center}
\includegraphics[width=17cm]{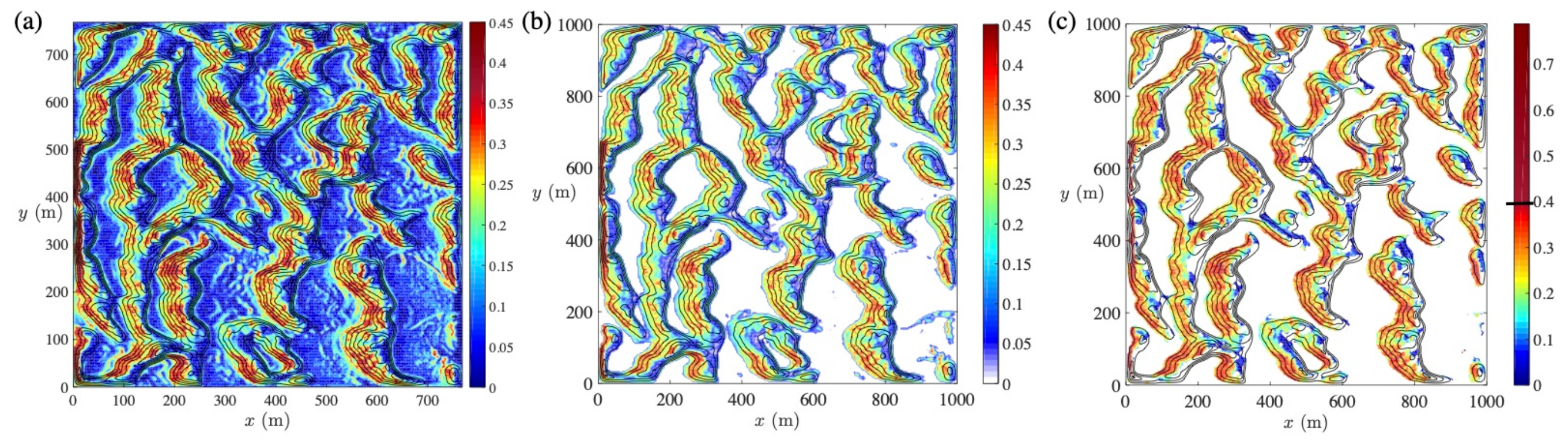}
\caption{Surface shear stress magnitude are captured through Immersed Boundary Method (IBM) \ref{IBM}. Panel (a) displays the original surface shear stress distribution. Plane (b) is surface shear stress on dunes which excludes the values on walls. Based on Equation \ref{facets}, the author prepared Panel (c) which exhibits the surface shear stress on windward face. Solid black line on colorbar of Panel (c) indicates the average value of high stress region.}
\label{Figure2}
\end{center}
\end{figure}

In this work, several mathematical models have been proposed to capture the real dune field stress. To achieve that goal, the aeolian stress distribution over dune field should be understood first. Previously, \citet{AndersonChamecki14} has revealed the Kelvin-Helmholtz effect in White Sands National Monument dune field, where the spanwise vortex rollers shed from the previous dune crestline and propagate in downwelling. The structural model of mixing layer flow in aeolian crescentic dune field sublayer has been proposed in \cite{AndersonChamecki14}, wherein the shear length $L_s$ is observed though LES results, which is in half length scale of the vortex thickness, $L_s=0.5\delta_\omega$. \citet{wang2019turbulence} has revealed the mixing-layer analogy in WSNM, where the vortex thickness is scaled with interdune mixing layer eddies, $\delta_\omega \sim L_{\omega}$. This is important, because it helps us to understand the role of turbulence played in this problem. That is the secondary flow induced by streamwise obstructive effects enhances the aeolian sediment erosion. Thus, hight scaled turbulence correlation associated with logarithmic upcoming wind profile unavoidably makes local wall-normal elevation become a significant variable to predict surface stress. However, the nonlinear correlation between vertical elevation and second order statistics induces several following mathematic models to capture surface stresses. Due to the low stress magnitude in leeward faces, the following practices are focussing on stoss side stress analysis. 

From the numerical analysis of surface stress, the stress magnitude displays a pseudo-linear correlation with dune stoss side gradient. Given the realistic dune morphologies, $\beta$ has been adopted to present streamwise and transversal gradient effects, where
\begin{equation}
\beta({\boldsymbol{x}})=\sqrt{\left[{\frac{\partial{h}}{\partial{x}}}\right]^{2}+\left[{\frac{\partial{h}}{\partial{y}}}\right]^{2}} S(\boldsymbol{x}).
\label{beta}
\end{equation}
$S(\boldsymbol{x})$ is defined in Eq.\ref{Sx} to filter out the leeward area. Here, a function $\gamma_n(\boldsymbol{x})$ is used to represent upcoming wind impact,
\begin{equation}
\beta_n(x,y)=\beta(x,y)\gamma_n(x,y).
\label{beta_new}
\end{equation}
To assess the reliability of each model, a residual factor is defined as following,
\begin{equation}
\xi=|\beta-\tau_w|,
\label{epsilon}
\end{equation}
where $\tau_w$ is the LES results. To simplify the assessing progress, the horizontal plane averaged residual value is also used here to represent the general performance of each model,
\begin{equation}
\xi_t=\frac{\Sigma^{Lx}_{0}\Sigma^{Ly}_{0}|\xi(x,y)|}{L_x\cdot L_y} = \frac{\Sigma^{Lx}_{0}\Sigma^{Ly}_{0}|\beta(x,y))-\tau_w(x,y)|}{L_x\cdot L_y}.
\label{xi_t}
\end{equation}

Table 1 displayed all mathematic models proposed in this work. The first model $\beta_0$ only considered dune stoss gradient effect. Assuming the linear wind profile, $\beta_1$ is proposed to certify the significance of aeolian process in sediment saltation. $\beta_2$ and $\beta_3$ includes the best fitting of wind profiles in trigonometric function formats. Considering the realistic boundary layer structures, $\beta_4$ to $\beta_7$ consider the upcoming velocity profile gradient.
\begin{table*}  %
\caption{Dune surface shear stress mathematic model $\beta$ and average residual value, where $h(\boldsymbol{x})$ is dune field wall-normal elevation, $L_s$ is shear length $L_s\approx h_{max}$, $h_a$ is inflection height $h_a=3h_{max}/4$, $h_{max}$ is the maximum value of dune height.}
\begin{center}
\begin{tabular}{|p{3cm}  p{5cm} p{2cm} |}
\hline
$\beta_n$							& $\gamma_n$																	& $\xi_{t}$ (Eq\ref{xi_t}) 	\\
\hline
$\beta_0=\beta$					& $\gamma_0=1.0 $																	&0.20		\\		
$\beta_1=\beta \gamma_1$		& $\gamma_1=h(x,y)/h_{max}$														&0.15		\\
$\beta_2=\beta \gamma_2$		& $\gamma_2=sin(h \pi/2h_{max})$													&0.15		\\
$\beta_3=\beta \gamma_3$		& $\gamma_3=tan(h\pi/(2h_{max}\cdot k))$											&0.17		\\
$\beta_4=\beta \gamma_4$		& $\gamma_4=tan\left(\frac{h\pi}{2h_{max}}-k\right)+m$								&0.29		\\
$\beta_5=\beta \gamma_5$		& $\gamma_5=tanh\left[1-\left({\frac{\frac{h}{h_{max}}-h_a}{L_s}}\right)^{2}\right]$		&0.15		\\
$\beta_6=\beta \gamma_6$		& $\gamma_6=1-tanh^2\left(\frac{h-h_a}{L_s}\right)$									&0.17		\\
$\beta_7=\beta \gamma_7$		& $\gamma_7=1+tanh\left(\frac{h-h_a}{L_s}\right)$										&0.15		\\
\hline
\end{tabular}
\end{center}
\label{Table2}
\end{table*}

\begin{figure}
\begin{center}
\includegraphics[width=16.5cm]{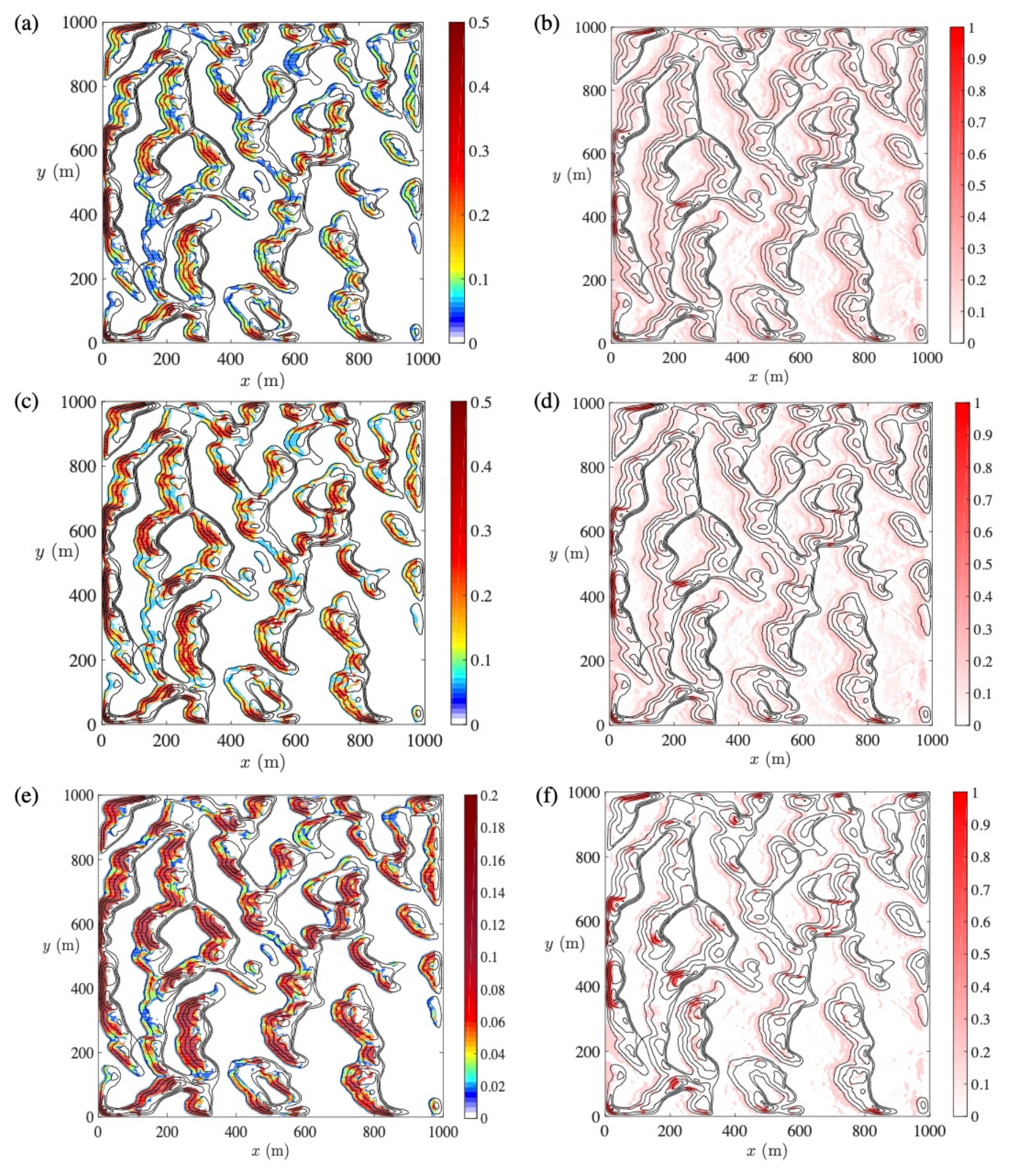}
\caption{The author prepared the value drawn from different mathematic models: Pane (a,b) show $\beta_1(\boldsymbol{x})$ and $\xi_1(\boldsymbol{x})$; Pane (c,d) show $\beta_2(\boldsymbol{x})$ and $\xi_2(\boldsymbol{x})$; Pane (e,f) show $\beta_3(\boldsymbol{x})$ and $\xi_3(\boldsymbol{x})$ respectively.}
\label{Figure3}
\end{center}
\end{figure}

\begin{figure}
\begin{center}
\includegraphics[width=16.5cm]{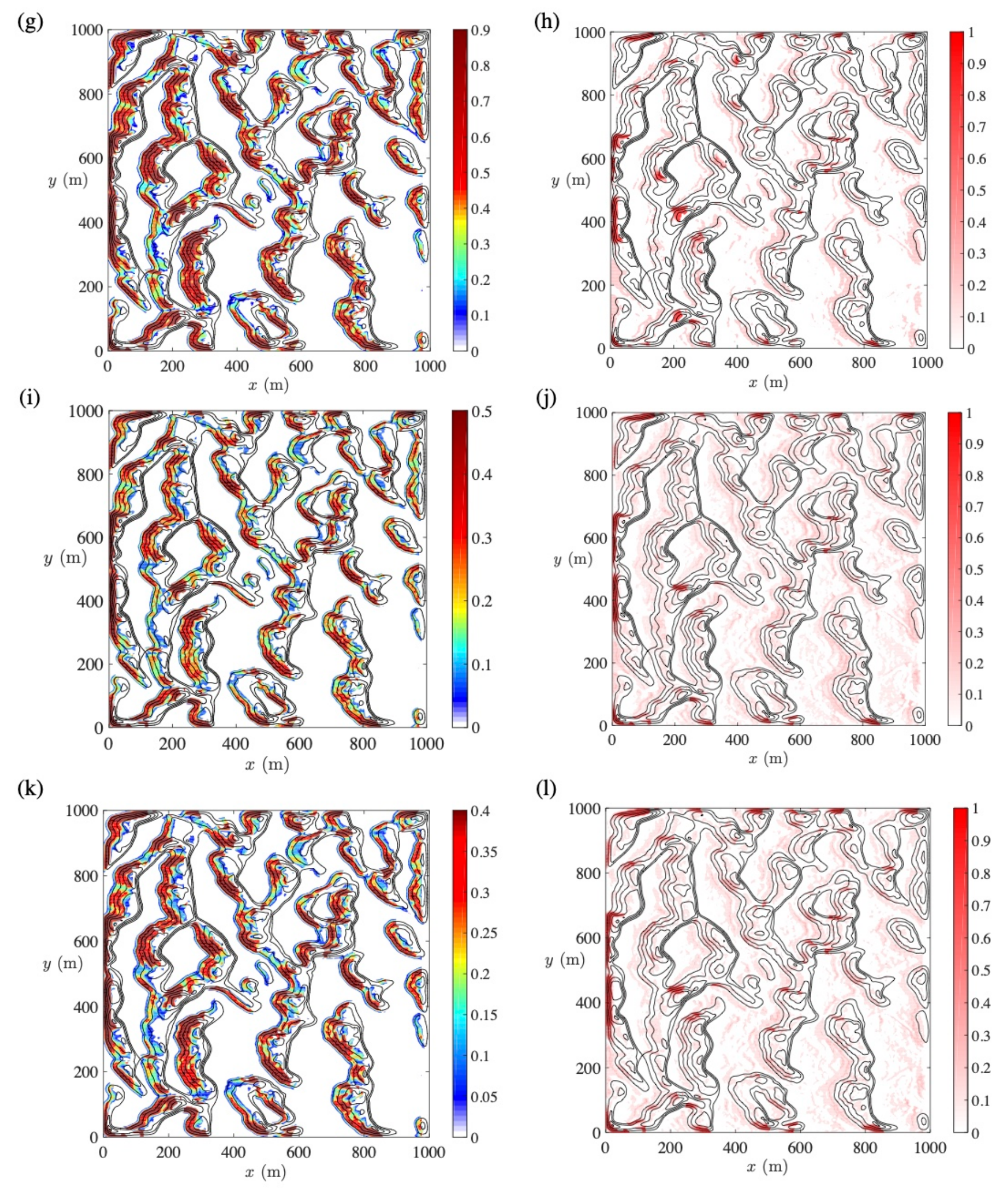}
\caption{The author prepared the value drawn from different mathematic models: Pane (g,h) show $\beta_4(\boldsymbol{x})$ and $\xi_4(\boldsymbol{x})$; Pane (i,j) show $\beta_5(\boldsymbol{x})$ and $\xi_5(\boldsymbol{x})$; Pane (k,l) show $\beta_6(\boldsymbol{x})$ and $\xi_6(\boldsymbol{x})$ respectively.}
\label{Figure4}
\end{center}
\end{figure}

\begin{figure}
\begin{center}
\includegraphics[width=16.5cm]{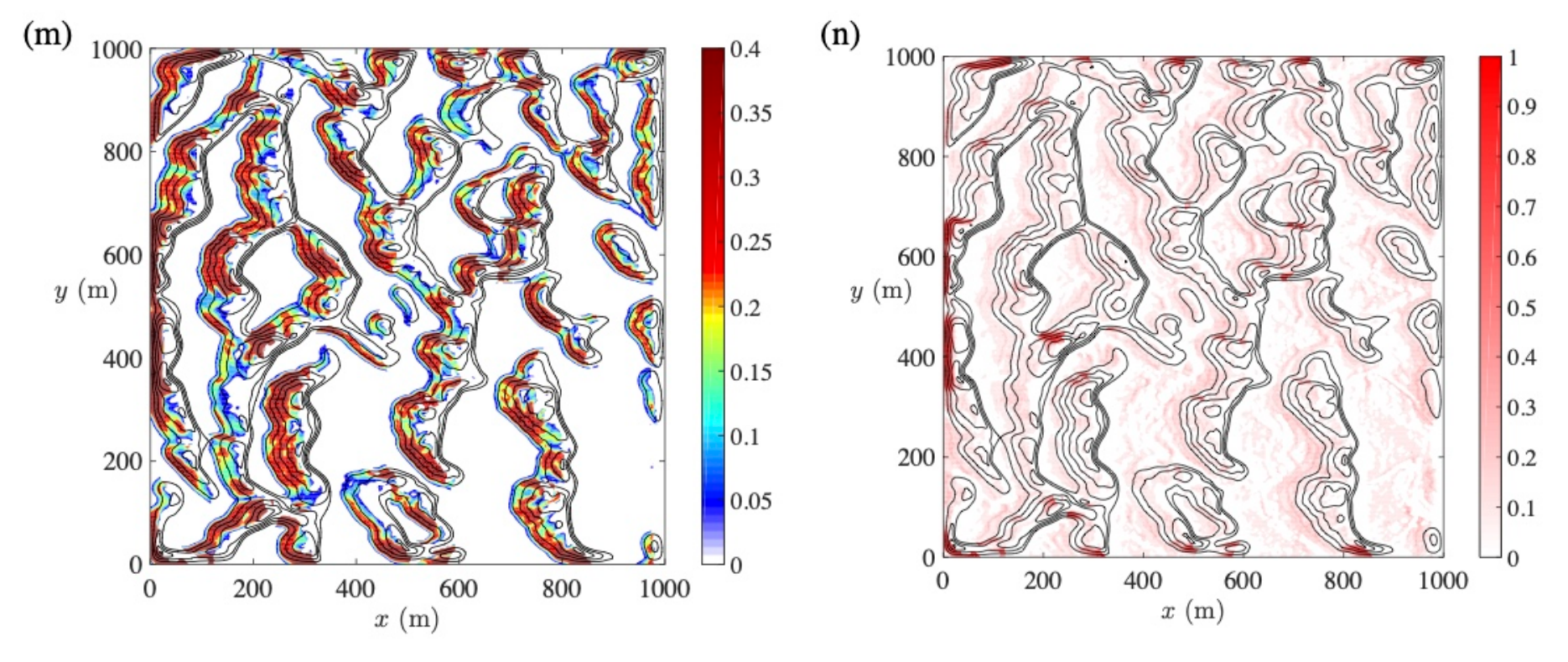}
\caption{The author prepared the value drawn from different mathematic models: Pane (m,n) show $\beta_7(\boldsymbol{x})$ and $\xi_7(\boldsymbol{x})$ respectively.}
\label{Figure5}
\end{center}
\end{figure}

Figure 3, 4 and 5 show the stress model alone with corresponding residual value distributions. Meanwhile, the average residual $\xi_t$ results are displayed in Table 1. The mathematic model results in Figure 3 to 5 are all showing reasonable stress magnitude and distribution. The residual distribution (f), (h) indicate $\beta_3$, $\beta_4$ and $\beta_5$ have the best performances. From $\xi_t$ in Table 1, the minimum value $min(\xi_t)$ appears in $\beta_1$, $\beta_2$, $\beta_5$ and $\beta_7$. Thus, based on the comprehensive assessment of residual value, $\beta_3$ and $\beta_5$ are preforming the best evaluation. 

In $\beta_3$ model, the mathematic model is
\begin{equation}
\beta_3(\boldsymbol{x})=\beta(\boldsymbol{x})\gamma_3(\boldsymbol{x})=\sqrt{\left[{\frac{\partial{h}}{\partial{x}}}\right]^{2}+\left[{\frac{\partial{h}}{\partial{y}}}\right]^{2}} S(\boldsymbol{x})tan(h\pi/(2h_{max}\cdot k)),
\end{equation}
where $\gamma_3$ is the best fitting of local wind profile. In $\beta_5$, the model is 
\begin{equation}
\beta_5(\boldsymbol{x})=\beta(\boldsymbol{x})\gamma_5(\boldsymbol{x})=\sqrt{\left[{\frac{\partial{h}}{\partial{x}}}\right]^{2}+\left[{\frac{\partial{h}}{\partial{y}}}\right]^{2}} S(\boldsymbol{x})tanh\left[1-\left({\frac{\frac{h}{h_{max}}-h_a}{L_s}}\right)^{2}\right].
\end{equation}

The profile of $\gamma_5$ is the gradient of theoretical boundary layer profile over dune field. According to \cite{wang2019turbulence,AndersonChamecki14}, boundary layer structure over dune field is composed of two regions: inertial sublayer and mixing layer. The velocity profile within inertial layer, $z>\delta_\omega$, is 
\begin{equation}
\frac{U(z)}{u_{*,d}}=\frac{1}{\kappa}ln\left[{\frac{z}{z_0}}\right],
\label{loglaw}
\end{equation}
where in mixing layer region, $0\leqslant z \leqslant\delta_\omega$ \cite{wang2019thesis,wang2020turbulent,michalke64,raupach96,metias96,katuletal},
\begin{equation}
\frac{U(z)}{u_{*,d}}=\frac{U_0}{u_{*,d}}\left[{1+tanh\left(\frac{z-h_a}{L_s}\right)}\right],
\label{mixlayer}
\end{equation}
\citet{AndersonChamecki14} has shown the elevated mean flow gradient in the roughness sublayer is responsible for the enhanced downward transport of high momentum fluid via turbulent sweep events. The gradient of velocity profile within inertial sublayer over dune field should be 
\begin{equation}
\frac{dU(z)}{dz}=\frac{u_{*,d}}{\kappa z},
\label{loggradient}
\end{equation}
but in within mixing sublayer
\begin{equation}
\frac{dU(z)}{dz}=\frac{U_0}{L_s}\left[1-\left({\frac{z-h_a}{L_s}}\right)^{2}\right].
\label{gradient2}
\end{equation}
Figure 6 shows the vertical profiles of averaged streamwise velocity and velocity gradient. In Panel (a), beneath the dotted red line, solid black line can collapse with dashed black line, which is the vertical profile of Eq.\ref{mixlayer}. However, at the very high elevation, solid black line can match solid blue line, which is Eq.\ref{loglaw}. In Panel (b), the mixing layer model is still validate for velocity gradient beneath the inflection height. And at higher elevation, the flow is within inertial layer which should obey the conical logarithmic gradient profile Eq.\ref{loggradient}. Figure 6 shows vertical profile of velocity gradient and the vertical profile of wind velocity, evaluated numerically. The $tanh$ function has closely captured the sublayer gradient form. That the maximum gradients match is simply a result of using the velocity profiles to evaluate $h_a$ and $L_s$, but the profiles also share a similar form. Similarly, we see the WSNM gradient tends towards the Eq.\ref{loggradient} profile for $z\gtrsim\delta_\omega$. These results of the dune field sublayer are consistent with the underlying obstructed shear flow arguments \cite{AndersonChamecki14,ghisalberti09,wang2020turbulent,bristow2019spatial}. 

\begin{figure}
\begin{center}
\includegraphics[width=10cm]{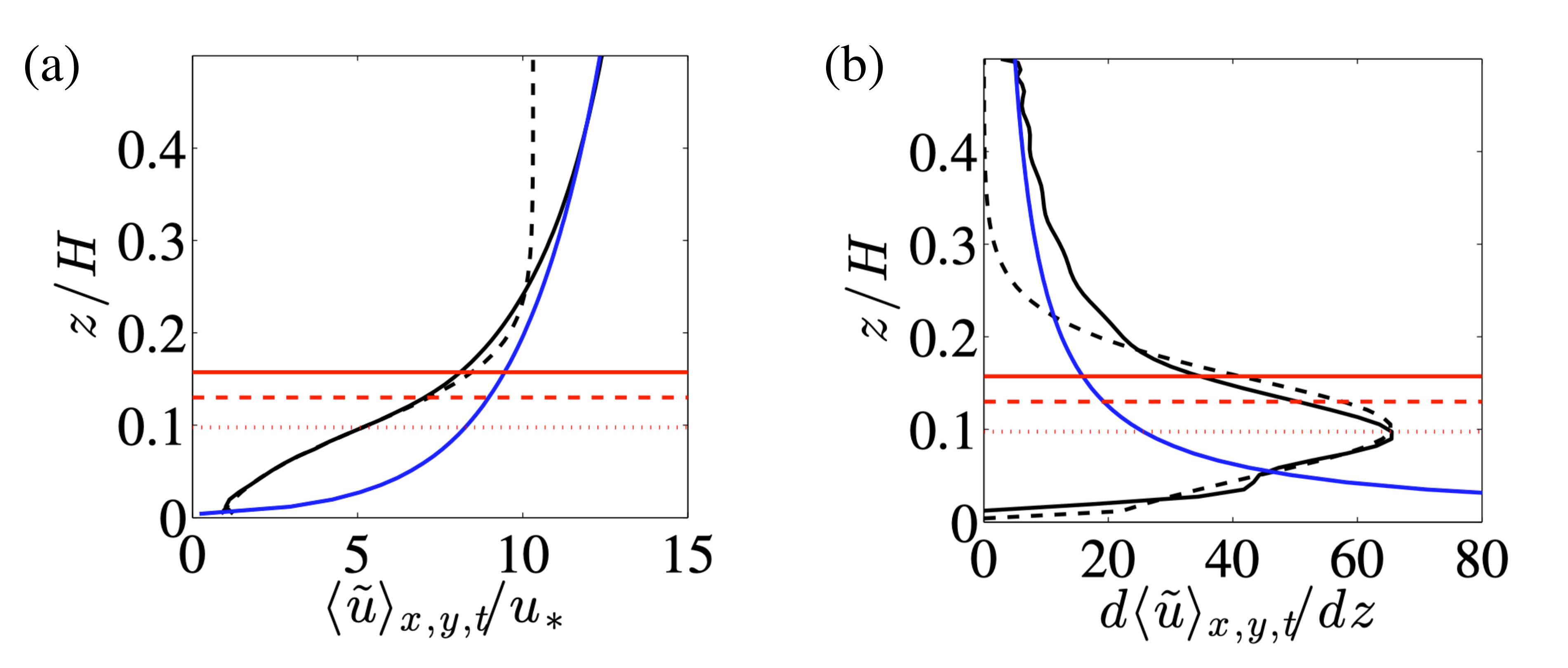}
\caption{Panel (a), solid black line is wall-normal plane-averaged streamwise velocity of WSNM (solid black line). Dashed black line shows profile in Eq.\ref{mixlayer}. Solid blue line is the profile of Eq.\ref{loglaw}. Panel (b), solid black line is vertical profile plane-averaged streamwise velocity gradient over WSNM. Dashed line shows profile in Eq.\ref{gradient2}. Solid blue line indicates the gradient profile of logarithmic wind velocity profile. Solid red line indicates the height of $\delta_\omega$. Dashed red line indicates the height of crest height $h_w$. Dotted red line indicates inflection height $h_a$.}
\label{Figure6}
\end{center}
\end{figure}

From generally global residual distributions, the locations with critical residual values show an evident consistency. These locations are defined as ``channeling'' regions \cite{wang2017numerical,bristow2017experimental,wang2017large,wang2018large,wang2018large2,wang2019turbulence,wang2019numerical2,wang2019thesis}. Two types of perturbation are triggered due to local dune obstructions, which are illustrated through Figure 7. The first type is named ``sediment scour'' (Figure 7 (a)), which is spanwise rotating secondary flow keeps scouring the sediment on inner dune faces \cite{wang2019thesis,wang2018large}. The second type is ``flow channeling'' \cite{wang2017numerical}. It indicates the enhancement of velocity magnitude in the narrow interdune region. Previously, \citet{wang2019turbulence} revealed the coherent structures in WSNM. That is streamwise and spanwise vortex rollers. Within mixing layer region, $z<\delta_\omega$, Kelvin-Helmholtz instability is featured in dune vortex shedding from brinklines. Spanwise vortex rollers is scaled with mixing layer length scales. Within the interdune regions, Prandtl's secondary flow of the first and second kind keeps feeding the interdune roller evolution \cite{wang2018large}, wherein turbulent coherency is broken down into aggregations of small length scale rollers, as relative spacing decreases. 
\begin{figure}
\begin{center}
\includegraphics[width=16.5cm]{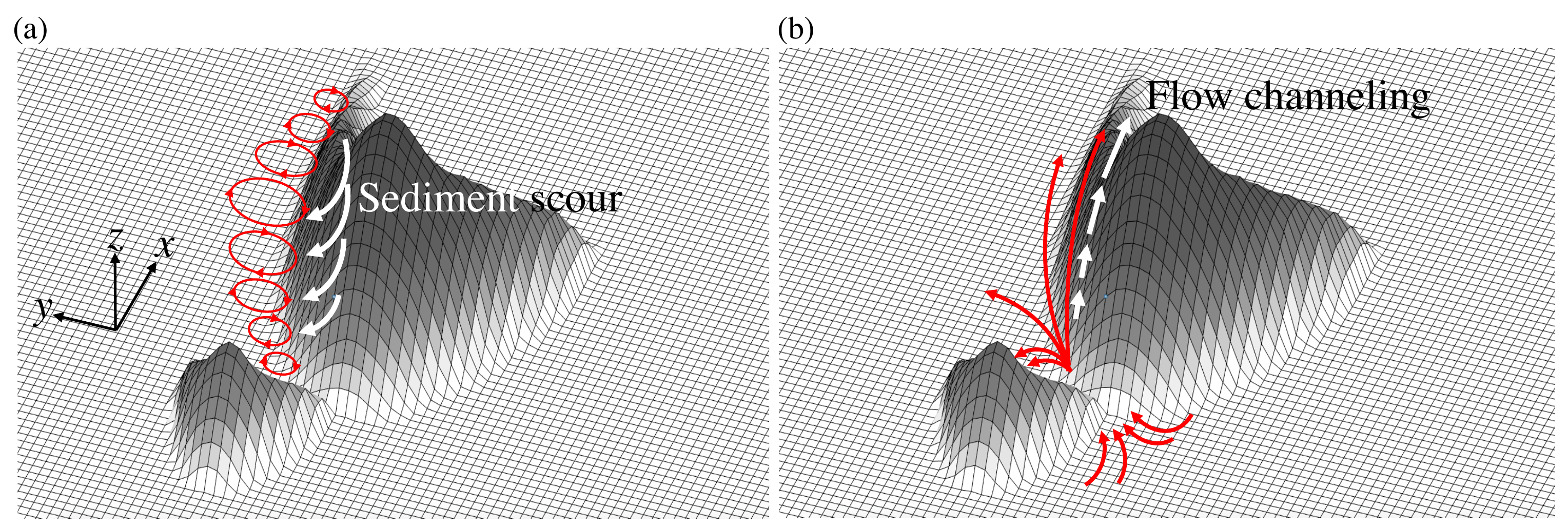}
\caption{The schematic drawing is displayed here which has been previously published in \citep{wang2018large}. Panel (a) shows sediment scour, where red quivers indicate the rotating direction in the flow channeling region, and white quivers show sediments eroded to ground by sediment scourring. Panel (b) exhibits flow channeling effect in interdune region. Due to mass conservation, the obstructive effect of downwind dune will increase the interdune flow velocity and surface shear \cite{wang2017numerical}. The white quivers indicate the sediment moving direction. Flow channeling enhanced surface shear triggers larger amount of sediment erosion \citep{wang2019thesis}.}
\label{Figure7}
\end{center}
\end{figure}

\section*{Acknowledgement}
The author acknowledge Dr. William Anderson (UT Dallas) for his guidance of this work, and Dr. Gary Kocurek and Dr. David Mohrig (University of Texas at Austin) for graciously providing the LiDAR survey of the White Sands National Monument (with support from the National Park Service and the National Science Foundation), and Dr. Ken Christensen (University of Notre Dame) for sharing idealized dune DEMs. Computational resources were provided by Texas Advanced Computing Center (TACC) at the University of Texas at Austin.

\bibliography{mybib.bib}  

\end{document}